\documentclass[twocolumn,showpacs,preprintnumbers,amsmath,amssymb]{revtex4}

\usepackage{graphicx}
\usepackage{dcolumn}
\usepackage{bm}

\begin{document}

\title{Dark Energy in the Universe, the Irreversibility of Time and Neutrinos}

\author{N. E. Mavromatos}

\email{Nikolaos.Mavromatos@kcl.ac.uk}

\affiliation{King's College London, Department of Physics, Theoretical
Physics, Strand, London WC2R 2LS, U.K.} 

\date{\today}

\begin{abstract}
In this review, I 
discuss briefly how the presence of a cosmological constant in the 
Universe may imply a decoherent evolution of quantum matter in it,
and as a consequence a fundamental irreversibility of time 
unrelated in principle to CP properties (Cosmological CPT Violation).
In this context, I also discuss recently suggested 
novel possible contributions of 
massive neutrinos to the cosmological constant, which are not due  
to the standard loop expansion in quantum field theory, 
but rather due to unconventional
properties of (some version of) the quantum theory underlying flavour mixing.
It is also argued that quantum space time foam may be responsible for the 
neutrino mass differences, observed today, 
and through the above considerations, 
for the (majority of the) dark energy of the Universe in the present era. 
In the above context, I also present 
a fit of all the currently available neutrino oscillation 
data, including the LSND ``anomalous'' 
experimental results, based on such a CPT Violating decoherent
neutrino model. The key feature is to use different decoherent
parameters between neutrinos and antineutrinos, due to the 
above-mentioned CPT violation. This points to 
the necessity of 
future experiments, concentrating on the antineutrino sector, 
in order to falsify the model. 

\end{abstract}

\maketitle

\section{Introduction}

Recent astrophysical observations, using different experiments
and diverse techniques, seem to indicate that 70\% of the 
Universe energy budget is occupied by ``vacuum'' energy density of 
unknown origin, termed Dark Energy~\cite{snIa,wmap}. 
Best fit models give the positive cosmological {\it constant} Einstein-Friedman
Universe as a good candidate to explain these observations, although
models with relaxing to zero vacuum energy (quintessential, i.e. 
involving a scalar field which has not yet 
reached the minimum of its potential) are 
compatible with the current data.

From a theoretical point of view the two categories of Dark Energy models 
are quite different. If there is a relaxing to zero cosmological vacuum energy,
depending on the details of the relaxation rate, it is possible in general
to define asymptotic states and hence a proper Scattering matrix (S-matrix) 
for the theory, which can thus be quantised canonically. 
On the other hand, Universes with a 
cosmological {\it constant} $\Lambda > 0$  (de Sitter) 
admit no asymptotic states, as a result of the Hubble horizon which 
characterises these models, and hampers the definition of proper asymptotic
state vectors, and hence a proper S-matrix. 
Indeed, de Sitter Universes will expand for ever, and eventually
their constant vacuum energy density component will dominate
over matter in such a way that the Universe will enter again an 
exponential (inflationary) phase of (eternal) accelerated expansion,
with a Hubble horizon of radius $\delta_H \propto 1/\sqrt{\Lambda}$. 
It seems that the recent astrophysical observations~\cite{snIa,wmap} 
seem to indicate
that the current era of the universe is the beginning of such an 
accelerated expansion.

Canonical quantisation of field theories in de Sitter space times is 
still an elusive subject, mostly due to 
the above-mentioned problem of a proper S-matrix
definition.
One suggestion towards the quantisation of such systems could be 
through analogies 
with open systems in quantum mechanics, interacting with an environment. 
The environment in cosmological constant models would consist of field modes
whose wavelength is shorter than 
the Hubble horizon radius. 
This splitting was originally suggested by Starobinski~\cite{staro},
in the context of his stochastic inflationary model, and later on was adopted
by several groups~\cite{coarse}. 
Crossing the horizon in either direction
would constitute interactions with the environment. 
An initially pure quantum state in such Universes/open-systems would 
therefore become eventually mixed, as a result of interactions 
with the environmental modes, whose strength will be controlled by the 
size of the Hubble horizon, and hence the cosmological constant. 
In particular, for some simple cases, 
such as conformally coupled scalar fields~\cite{coarse} 
in de Sitter spaces it has been shown explicitly that the 
system modes decohere if they have wavelengths longer than
a critical value, which is of the order of the Hubble horizon. 
Such decoherent evolution could explain the classicality of the early
(or late, in the case of a cosmological constant) Universe 
phase transitions~\cite{rivers}. The approach is still far from being 
complete, not only due to the technical complications, 
which force the researchers to adopt severe, and often unphysical,
approximations, but also due to conceptual issues, most of which
are associated with the back reaction of matter onto space time,
an issue often ignored in such a context. It is my opinion
that the latter issue plays an important r\^ole in the evolution
of a quantum Universe, especially one 
with a cosmological constant, and is associated
with quantum gravity issues. The very origin of the cosmological constant,
or in general the dark energy of the vacuum, is certainly a property
of quantum gravity.

Since string theory seems to be the most rigorous 
and most successful 
approach to quantum gravity, to date, 
encompassing the known quantum field theories of flat space times
in its low energy limit, 
I would like to approach 
the  problem of the cosmological constant in this framework. 
Critical string theory models, 
which are based on S-matrix theory, at least for their
perturbative formulation, can only accommodate relaxing to 
zero vacuum energy models, which allow for a proper definition
of asymptotic states, but cannot deal 
with de Sitter Universes~\cite{stringds}. On the other hand, 
string/membrane theory models with anti-de-Sitter 
backgrounds, which admit supersymmetry, are consistent.
In critical string theory an evolution 
of pure states to mixed do not exist,
and this is another way 
of understanding the incompatibility of conventional strings 
with de Sitter Universes. However, quantum effects in strings,
generated by dilaton tadpoles in $\sigma$-models formulated in 
world sheets with  higher topologies (genera)
lead to non-zero contributions to 
the cosmological constant (Fischler-Susskind mechanism)~\cite{fs}.
Modular divergences in such theories require regularisation,
which essentially means that small handles in closed strings 
(appropriate for gravity), 
of size smaller than the world-sheet short-distance cutoff, 
must be integrated out in an effective Wilsonian
path integral, since such small handles could not be distinguished
from tree-level world-sheet topologies.
This corresponds to adding to the tree level $\sigma$-model 
action a world-sheet counter-term appropriate for absorbing such 
modular (small handle) divergences. The effects of such terms 
lead to corrections in the 
tree-level graviton world-sheet $\beta$-functions similar to those arising 
in a de Sitter Universe. However, the issue of the impossibility 
of defining a proper S-matrix in de Sitter Universe brings up
an immediate question on the consistency of the approach
within string theory.

\section{Non-Critical Strings in de Sitter Backgrounds} 

Fortunately there is a way out, which goes beyond the 
above-described Fischler-Susskind
mechanism for generating cosmological constant contributions.  
From a world-sheet view point, a non-zero cosmological 
constant amounts to 
contributions to the effective central charge of the two-dimensional 
world-sheet field theory, which thus deviates from its (conformal point) 
critical 
value.
Stringy $\sigma$-models with a non-zero central charge deficit
constitute the so-called Liouville strings~\cite{ddk}. 
The path-integral world-sheet quantisation of such non-critical
strings requires the introduction of the Liouville mode $\phi$, 
which is an extra world-sheet field whose target-space signature 
(time-like or space-like) depends on the signature of the central charge
deficit (positive or negative respectively). 
This extra field, plays the r\^ole of a new coordinate in target space,
and its presence is responsible for the restoration of the 
conformal invariance of the theory~\cite{ddk}.

An important step towards the physical significance of the 
Liouville mode is the identification~\cite{emn} 
of its world-sheet zero mode $\phi_0$ 
with the target time,
in the supercritical (positive central charge deficit) theories. 
Such an identification emerges from dynamics of the target space 
low-energy effective field theory~\cite{grav}, in the sense of 
minimisation of the effective potential. 
Furthermore it can be shown rigorously~\cite{emn} 
that under such an identification one cannot define a pure-state S-matrix
but rather a \$ -matrix, which is a non-factorisable product of $S$ and 
$S^\dagger $, acting on density matrix mixed states rather than pure states.
The non factorisability may be attributed to divergences in 
the short-distance
world-sheet behaviour of the $\sigma$-model theory. 

With this identification in mind, one may proceed to discuss the 
issue of propagation of quantum matter in a de Sitter Background, 
within such a non-critical
string framework. By following simple arguments on world-sheet 
renormalisation-group invariance
of $\sigma$-model quantities which have target-space physical relevance,
it is straightforward to arrive at the following master equation
describing the evolution of string low-energy matter 
in a non-conformal $\sigma$-model background:
\begin{equation} 
{\dot \rho} = i[\rho, H] + :\beta^i{\cal G}_{ij}[g^j, \rho]:
\label{liouveq}
\end{equation}
where 
$\rho$ is the density matrix of string matter excitations,
$H$ is the effective low-energy matter Hamiltonian,
$g^i$ are background target-space fields, and $\beta^i$ are their
corresponding $\sigma$-model renormalisation group $\beta$-functions,
expressing their scaling under Liouville dressing~\cite{ddk}.
Canonical quantisation for the operators/fields $g^i$ 
is possible in Liouville strings~\cite{emn}, 
as a result of summing up higher world-sheet topologies,
and thus 
$: \dots :$ in (\ref{liouveq}) denotes appropriate quantum ordering.  
The quantity ${\cal G}_{ij} = 2z^2{\overline z}^2<V_i(z)V_j(0)>$
is the so-called Zamolodchikov ``metric''
in the moduli space of the string, 
a two-point correlation function with respect to the vertex operators
$V_i$ corresponding to the deformations of the $\sigma$-model
action from the conformal point $S^*$: 
\begin{equation} 
S_\sigma = S^* = g^i\int_\Sigma V_i~,
\label{deform}
\end{equation} 
where $\int_\Sigma$ denotes integration over the world-sheet.  
The dot over $\rho$ in (\ref{liouveq}) 
denotes differentiation with respect to 
the world-sheet zero mode 
of the Liouville field, identified in this approach with the target time~\cite{emn}. An important note for the compact notation in (\ref{deform})
is now in order. The index $i$ runs over {\it both} species of background 
target space fields as well as space-time coordinates. 
Thus the summation over 
$i,j$ indices in (\ref{liouveq}) corresponds to a summation over $M,N$ indices
but also a continuous generally covariant 
space-time integration $\int d^dy \sqrt{-g}$,
where $y$ denotes a set of d-dimensional space-time coordinates. 
It is important
to stress that strings respect general covariance by construction. 
For instance, 
for the case at hand, where we are interested in perturbations of the metric 
background $g_{MN}$, where $M,N$ are target space-time indices, 
one has the correspondence: 
\begin{eqnarray} 
g^i &\to& g_{MN}(y)~, \nonumber \\
V_i &\to& V^{MN} (X,y)= \partial_\alpha  X^M  \partial^\alpha X^N \delta^{(d)} 
(y - X(\sigma,\tau))~, \nonumber \\
{\rm where} ~\alpha &=& \sigma, \tau~, \nonumber \\
\int _\Sigma g^i V_i &\to& \int d^dy \sqrt{-g} g_{MN}(y)
\int_\Sigma V^{MN}(X,y)~, 
\label{corr}
\end{eqnarray}

For conformal world-sheet 
backgrounds $\beta^i =0$ and one obtains a normal
quantum mechanical equation where purity of states is preserved under
evolution. When non-conformal string backgrounds are 
present, however, one has non-quantum mechanical corrections terms
in this evolution, which in general may imply decoherence of matter,
that is evolution of initially pure states to mixed ones. 
In a perturbative derivative expansion (in powers of 
$\alpha'$, where $\alpha' =\ell_s^2$ 
is the Regge slope of the string, and $\ell_s$ the fundamental string length),
the lowest order graviton $\beta$ function is just the Ricci tensor
\begin{equation}
\beta_{MN} = \alpha ' R_{MN} + \dots
\label{ricci}
\end{equation}
where the $\dots$ indicate terms higher order in $\alpha '$, which can be ignored
in a low-energy (infrared) framework for the target space effective 
field theory, we are interested in here. From now on, unless 
otherwise stated, we shall work in units of $\alpha ' = 1$ for brevity.

Conformal backgrounds in string theory are therefore Ricci flat backgrounds
in this framework. On the other hand, de Sitter backgrounds, for which 
$R_{MN} \propto \Lambda g_{MN} \ne 0$, with $\Lambda > 0$ a cosmological 
constant, obviously violate this condition, and the excitation 
of strings in such backgrounds can be described, at least perturbatively
for the (physically relevant) case of 
small $\Lambda$, so that the deviation from the 
Ricci flatness is minute, 
by means of a non-critical Liouville dressed $\sigma$-model.
From (\ref{liouveq}), (\ref{corr}), 
the evolution equation of low-energy string matter in such a background, then,
reads in this case~\cite{mlambda}:
\begin{equation} 
{\dot \rho} = i[\rho, H] + \int d^dy \Lambda : \sqrt{-g}g_{MN}[g^{MN}, \rho]:
\label{liouveq2}
\end{equation} 
where we took into account the fact that to lowest order, the 
Zamolodchikov metric is just the appropriate tensorial identity in 
all sets of indices (M  and $y$). If one chooses an antisymmetric
ordering prescription, and adopts a weak-graviton expansion
about flat Minkowski space time, $g_{MN} = \eta_{MN} + h_{MN}$,  
which seems to be the cosmologically relevant case for the present 
era of the Universe,
then one arrives at a double commutator
structure for the decoherence term~\cite{mlambda}:
\begin{equation} 
{\dot \rho} = i[\rho, H] + \int d^dy \Lambda [h_{MN}, [h^{MN}, \rho]]
\label{liouveq3}
\end{equation} 
Notice that the decoherence term, which is real,
is not invariant, due to its structure, 
under the 
time reversal symmetry $t \to -t$ (we remind the reader that 
under such a symmetry, the matter Hamiltonian is time reversal invariant, 
but the $i \to -i$. Moreover, since we are dealing with 
small perturbations around flat Minkowski space time, the quanta
of the gravitational field $h_{MN}$ can be taken to respect the 
time reversal symmetry).  This breaking of the time reversal invariance
is unrelated in principle to 
properties of matter under the discrete symmetries of 
Charge (C) and Parity (P), and thus the $\Lambda$-induced 
decoherence term is CPT violating. 
This is what I would call
Cosmological CPT Violation~\cite{mlambda}, due to 
the global nature of the non-quantum mechanical terms in (\ref{liouveq3}).
Moreover, taking into account that $\Lambda g_{MN}$ in de Sitter
spaces may be viewed as a contribution to the stress tensor 
$T_{MN}^{\rm vac}$ of the vacuum, 
one observes that the decoherence term in (\ref{liouveq3}) 
may be considered as a quantum version of the (integrated over space time) 
trace of this tensor, 
thereby being proportional to 
the global conformal anomaly of the de Sitter space-time 
vacuum. 

The above results are in full agreement with the violation of CPT in 
decoherent field theories  characterised by an 
evolution of pure to mixed quantum states, which we have here 
due to the presence of the Hubble horizon~\cite{coarse}.
Indeed, according to a mathematical theorem by R.~Wald~\cite{wald}, the 
CPT operator is not a well-defined quantum mechanical operator
in field theories where there is decoherence, that is evolution of 
pure to mixed states. This leads to a violation of CPT symmetry
in its strong form, or rather {\it microscopic time irreversibility}. 
This may lead to different decoherent parameters eventually between
particles and antiparticles, reflecting the different ways 
of interaction with the foam between the two sectors. 
We should remark here that, 
in the case of two-state systems, such as two generation
neutrino oscillation models, the double commutator terms proportional 
to the cosmological constant in (\ref{liouveq3}) 
may be expressed in terms of metric variations 
$(\Delta g_{MN})^2$  between, say, neutrino energy eigenstates,
expressing back reaction of neutrino fluctuations onto the space time,
as a consequence of interaction with the foam~\cite{bm}. 
The induced CPT violation may in general imply, then, 
metric variations of different strength 
between particle and antiparticle sectors.

On the other hand, as stated in \cite{wald}, it could be possible
that, despite the strong violation of CPT, a weaker form of 
CPT invariance is maintained phenomenologically, in the sense that
an observer can always prepare pure initial states $|\phi>$, 
which could evolve to 
pure final states, $|\psi>$,
and for this {\it subset of states} the probabilities
for the transition and its CPT image were equal:
\begin{equation}
P(\phi \to \psi) = P(\theta^{-1} \psi \to \theta \phi)
\label{equality} 
\end{equation} 
where $\theta $ is the anti-unitary CPT operator acting on pure states only.
Such an issue can 
be disentangled experimentally, and this is the next topic in our discussion.

\section{Checking Microscopic Time Irreversibility in the Lab: Neutrinos}

The most sensitive, and physically interesting, particle probe for 
quantum-gravity decoherence, to date, appears to be the neutrino,
for which recently there is mounting experimental evidence
that it carries a non-trivial mass. The inequality of neutrino masses
among the various flavours leads to oscillations, whose
properties are affected by the above-mentioned decoherent
evolution.

In general, decoherent evolution may be induced by other means,
such as the presence of ordinary matter, which the particle 
passes through, or the presence of quantum space-time foam situation,
in which microscopic (Planck size) topologically non-trivial
metric fluctuations may make the ground state of quantum gravity
behave as a `medium'. Such contributions are 
in general independent of the 
above-described cosmological CPT Violation, although, as we shall
discuss below, there might be a common origin 
of both quantum space-time foam decoherence and cosmological constant
in the following sense: according to some speculative scenaria~\cite{bm2}, 
a neutrino 
mass difference, and hence flavour mixing, 
might be 
the result of quantum gravity decoherence,
in analogy with the celebrated MSW effect~\cite{msw}, where 
contributions to the mass difference
between neutrino flavours is induced as a result of the passage
of neutrinos through ordinary media. 
In some approach to the quantisation of flavour mixing 
in field theory~\cite{vitiello1,vitiello2}, as we shall
discuss in the next section, one can show that there
are non-trivial non-perturbative contributions to the vacuum energy
of the Universe (cosmological constant) 
from massive neutrinos, which are in fact
proportional to the (sum) of the mass differences, at least in 
hierarchical neutrino models~\cite{bm2}. According to our 
discussion above, then, this would imply cosmological CPT Violation,
pointing to the inapplicability
of flat-space methods for the quantisation of the flavour space,
which is thus becoming a curved-space time (de Sitter) problem,
awaiting solution.

From a phenomenological viewpoint, one can adopt a 
model-independent approach to arrive at the master equation
for the time evolution of the neutrino density matrix,
which could encompass all such foam or cosmological constant 
CPT Violating effects
in a unified formalism, without the necessity for 
a detailed microscopic knowledge of the underlying physics
for the `environment'. For three generation neutrinos 
this has been done in \cite{bm}, and we next proceed to review
briefly the situation.  

The mathematical formalism adopted is the 
so-called Lindblad or mathematical semi-groups approach
to decoherence~\cite{lindblad},
which is a very efficient way of studying open systems
in quantum mechanics. The time irreversibility in the evolution
of such semigroups, which is linked to decoherence, 
is inherent in the mathematical property 
of the lack of an inverse in the semigroup. This approach has been 
followed for the study of quantum-gravity decoherence in the 
case of neutral kaons in \cite{ehns,lopez}.

The Lindblad approach to decoherence does not require any detailed knowledge
of the environment, apart from energy conservation, 
entropy increase and complete positivity of the (reduced) density 
matrix $\rho(t)$ of the subsystem under consideration. 
The basic evolution equation for the (reduced) density matrix of the 
subsystem in the Lindblad approach is {\it linear} in $\rho(t)$ and 
reads: 
\begin{equation} 
\frac{\partial \rho}{\partial t} = -i[H_{\rm eff}, \rho] + \frac{1}{2}\sum_{j}\left([b_j, \rho(t)b^\dagger_j] + [b_j\rho(t), b_j^\dagger]\right)~,
\label{lindblad}
\end{equation}
where $H_{\rm eff}$ is the effective Hamiltonian of the subsystem, 
and the operators $b_j$ represent the interaction with the environment,
and are assumed bounded. Notice that the Lindblad part
cannot be written as a commutator (of a Hamiltonian 
function) with $\rho$. Environmental 
contributions that 
can be cast in Hamiltonian evolution (commutator form) are absorbed
in $H_{\rm eff}$.

It must be noted at this stage that the requirement
of complete positivity, which essentially
pertains to the positivity of the map $\rho(t)$ as the time evolves
in the case of many particle situations, such as meson factories 
(two-kaon states ($\phi$-factory), or two-$B$-meson states {\it etc.}),
may not be an exact property of quantum gravity, whose interactions
with the environment could be {\it non linear}~\cite{emn}. 
Nevertheless, complete 
positivity leads to a convenient and simple parametrization, and 
it has been assumed so far in many phenomenological 
analyses of quantum gravity decoherence in generic two state systems,
such as two-flavor neutrino systems~\cite{lisi,benatti2,benatti}.

Formally, the bounded Lindblad operators of an $N$-level quantum 
mechanical system can be expanded in a basis of matrices 
satisfying standard commutation relations of Lie groups. 
For a two-level system~\cite{ehns,lopez} such matrices are the 
SU(2) generators (Pauli matrices) 
plus the $2\times 2$ identity operator, while for a three level
system~\cite{gago}, which will be relevant for our purposes in this article,
the basis comprises of  
the eight Gell-Mann $SU(3)$ matrices $\Lambda_i~,~i=1,\dots 8$
 plus the $3\times 3$ identity matrix $I_{3x3}$. 

Let ${\cal J}_\mu$, $\mu=0,\dots 8 (3)$ be a set of SU(3) (SU(2)) generators  
for a three(two)-level system; then, one may expand the various terms 
in (\ref{lindblad}) in terms of ${\cal J}_\mu $ to arrive at the generic form:
\begin{eqnarray}
&&\frac{\partial \rho_\mu}{\partial t} = \sum_{ij} h_i\rho_j {f}_{ij\mu}  
+ \sum_{\nu} {\cal L}_{\mu\nu}\rho_\mu~, \nonumber \\
&& \mu, \nu = 0, \dots N^2 -1, \quad i,j = 1, \dots N^2 -1
\label{expandedlind}
\end{eqnarray}
with $N=3 (2)$ for three(two) level systems, and $f_{ijk}$ the structure
constants of the $SU(N)$ group. 
The requirement for entropy increase implies the hermiticity of  the 
Lindblad operators $b_i$, as well as the fact that 
the matrix ${\cal L}$ of the the non-Hamiltonian part of the 
evolution has the properties that ${\cal L}_{0\mu}={\cal L}_{\mu 0} =0$,
${\cal L}_{ij} =\frac{1}{2}\sum_{k,\ell ,m} b_m^{(n)} b_k^{(n)}{f}_{imk}{f}_{\ell k j}$, with the notation $b_j \equiv \sum_\mu b_\mu^{(j)} {\cal J}_\mu $.   

In the two-level case 
of \cite{ehns} the decoherence matrix ${\cal L}_{\mu\nu}$ 
is parametrised by a $4\times 4$ matrix, whose non vanishing entries 
are occupied by the three parameters with the dimensions of energy 
$\alpha, \beta, \gamma$ with the properties mentioned above. 
If the requirement of a completely positive map $\rho (t)$ is imposed, 
then the $4 \times 4$ matrix ${\cal L}$ 
becomes diagonal, with only one non vanishing 
entry occupied by the decoherence parameter $\gamma > 0$~\cite{benatti}. 

In \cite{bm} the CPT Violation feature of space-time foam decoherence,
has been taken into account for the neutrino oscillation case, by assuming
{\it different } decoherence parameters between particle and antiparticle
sectors. Below we shall use the 
barred notation for the antiparticle sector quantities. 
Notice that this is possible in neutrino oscillations because
we are dealing with oscillations among flavours separately
between particle and antiparticle sectors, e.g. we shall be interested
in probabilities $P_{\nu_\alpha \to \nu_{\beta}}$, or 
$P_{\bar \nu_\alpha \to \bar \nu_{\beta}}$, where $\alpha, \beta$ are
neutrino flavours. In contrast, in neutral meson systems~\cite{ehns,lopez},
one is dealing with oscillations between particle antiparticle sectors
(e.g. $K^0 \to \bar K^0$ for Kaons, {\it etc.}), and hence the 
relevant decoherent evolution contains only one set of 
parameters ($\alpha,\beta,\gamma$) for both sectors. 

The extension of the completely
positive decoherence scenario to the standard three-generation neutrino 
oscillations case requires formally the adoption 
of the three-state Lindblad problem.
The relativistic neutrino Hamiltonian
$H_{\rm eff} \sim p^2 + m^2/2p$, with $m$ the neutrino mass,  
has been used as the Hamiltonian
of the subsystem in the evolution of eq.(\ref{lindblad}).
In terms of the generators ${\cal J}_\mu$, $\mu = 0, \dots 8$ 
of the SU(3) group, $H_{\rm eff}$ can be expanded as~\cite{gago}:
${\cal H}_{\rm eff} = \frac{1}{2p}\sqrt{2/3}\left(6p^2 + \sum_{i=1}^{3}m_i^2
\right){\cal J}_0 + \frac{1}{2p}(\Delta m_{12}^2){\cal J}_3
+ \frac{1}{2\sqrt{3} p}\left(\Delta m_{13}^2 + \Delta m_{23}^2 \right){\cal J}_8$, 
with the obvious notation $\Delta m_{ij}^2 = m_i^2 - m_j^2$, $i,j =1,2,3$.

The analysis of \cite{gago} assumed {\it ad hoc} a diagonal form 
for the $9 \times 9$ decoherence matrix ${\cal L}$ in (\ref{expandedlind}):
\begin{equation}
[{\cal L}_{\mu\nu}]= {\rm Diag}\left(0, -\gamma_1,-\gamma_2,-\gamma_3,-\gamma_4,-\gamma_5,-\gamma_6,-\gamma_7,-\gamma_8\right)
\label{diagonal}
\end{equation} 
in direct analogy with 
the two-level case of complete positivity~\cite{lisi,benatti}. 
As we have mentioned already, there is no strong physical
motivation behind such restricted forms of decoherence. This assumption,
however,
leads to  the simplest possible decoherence models, 
and, for our  
{\it phenomenological} purposes
in \cite{bm} and here,  
we will assume 
the above form and use it 
to fit all the available neutrino data. 
It must be clear to the reader though,
that such a simplification, if proven to be successful (which, as we shall 
argue below, is the case here),
just adds more in favour of decoherence models, 
given the 
restricted number of available parameters for the fit in this case. 
In fact, any other non-minimal
scenario will have it easier to accommodate data 
because it will have more degrees of freedom available for such a 
purpose. 

In this formalism, the neutrino transition probabilities
read~\cite{gago,bm}:
\begin{eqnarray} 
&& P(\nu_\alpha \to \nu_\beta) ={\rm Tr}[\rho^\alpha (t)\rho^\beta] 
= \nonumber \\
&&\frac{1}{3} + \frac{1}{2}\sum_{i,k,j}
e^{\lambda_k t}{\cal D}_{ik}{\cal D}_{kj}^{-1}\rho^\alpha_j (0)\rho_i^\beta
\label{trans}
\end{eqnarray}
where $\alpha,\beta = e, \mu, \tau$ stand for the three neutrino flavors, and 
Latin indices run over $1, \dots 8$. The quantities $\lambda_k$ 
are the eigenvalues of the matrix ${\cal M}$ appearing in the 
evolution (\ref{expandedlind}), after taking into account probability
conservation, which decouples $\rho_0(t)=\sqrt{2/3}$, leaving the remaining 
equations in the form: $\partial \rho_k /\partial t = \sum_{j} 
{\cal M}_{kj}\rho_j $. The matrices ${\cal D}_{ij}$ are the matrices
that diagonalise ${\cal M}$~\cite{lindblad}. Explicit forms of 
these matrices, the eigenvalues $\lambda_k$, 
and consequently the transition probabilities
(\ref{trans}), are given in \cite{gago}. 

The important point to stress is that, in generic models of oscillation plus 
decoherence, the eigenvalues $\lambda_k$ depend
on both the decoherence parameters $\gamma_i$ and the mass differences
$\Delta m^2_{ij}$. For instance, $\lambda_1 = \frac{1}{2}[-(\gamma_1 + \gamma_2)
-\sqrt{(\gamma_2 - \gamma_1)^2 -4\Delta_{12}^2}]$, with 
the notation $\Delta_{ij} \equiv \Delta m_{ij}^2/2p$, $i,j=1,2,3$.
Note that, to leading order in the (small) squared-mass differences, 
one may replace
$p$ by the total neutrino energy $E$, and this will be understood 
in what follows.
We note that 
$\lambda_k$ depend on  
the quantities 
$\Omega_{ij}$:
\begin{eqnarray}
\Omega_{12} &=& \sqrt{(\gamma_2 - \gamma_1)^2 -4\Delta_{12}^2} \nonumber \\
\Omega_{13} &=& \sqrt{(\gamma_5 - \gamma_4)^2 -4\Delta_{13}^2} \\
\Omega_{23} &=& \sqrt{(\gamma_7 - \gamma_6)^2 -4\Delta_{23}^2} \nonumber 
\end{eqnarray}
{}From the above expressions  
for the eigenvalues $\lambda_k $, it becomes clear
that, when decoherence and oscillations
are present simultaneously,
one should distinguish two cases, 
according to the relative magnitudes of $\Delta_{ij}$ and 
$\Delta\gamma_{kl} \equiv \gamma_k - \gamma_l$: (i) $2|\Delta_{ij}| \ge |\Delta\gamma_{k\ell}|$, and (ii) 
$2|\Delta_{ij}| < |\Delta\gamma_{k\ell}|$.
In the former case, the probabilities (\ref{trans}) contain trigonometric
(sine and cosine) functions, whilst in the latter they exhibit  
hyperbolic sin and cosine dependence.

Assuming mixing between the flavors, amounts to expressing neutrino
flavor eigenstates $|\nu_\alpha>$, $\alpha=e,\mu,\tau$ in terms of 
mass eigenstates $|\nu_i>$, $i=1,2,3$ through a (unitary) matrix $U$:
$|\nu_\alpha> = \sum_{i=1}^{3}U^*_{\alpha i}|\nu_i>$. This implies that
the density matrix of a flavor state $\rho^\alpha $ can be expressed 
in terms of mass eigenstates as: 
$\rho^\alpha=|\nu_\alpha><\nu_\alpha|=\sum_{i,j}U^*_{\alpha i}U_{\alpha j}|\nu_i><\nu_j| $. From this we can determine 
$\rho_\mu^\alpha =2{\rm Tr}(\rho^\alpha {\cal J}_\mu )$, a quantity needed to
calculate the transition probabilities (\ref{trans}).

Due to CPT Violation, as mentioned above, 
we should notice at this stage that, when considering
the above probabilities in the antineutrino
sector, the respective decoherence parameters ${\bar \gamma}_i$ 
in general may be different from the corresponding ones in the 
neutrino sector,
as a result of the strong form of CPT violation. 
This will be crucial for accommodating~\cite{bm} 
the LSND result~\cite{LSND} without conflicting with  
the rest of the available neutrino data.
This feature is totally unrelated to mass differences between flavors.

Compatibility of all available neutrino 
data, including CHOOZ~\cite{chooz} and LSND~\cite{LSND}, 
can be achieved 
through a  
set of decoherence parameters $\gamma_j$ 
in (\ref{diagonal}) such that: 
all the $\gamma_i$ in the neutrino sector are set to zero, 
restricting in this way 
all the decoherence effects to the antineutrino one where: 
\begin{eqnarray} 
 \overline{\gamma_{1}} = \overline{\gamma_{2}} = \overline{\gamma_{4}} &=& 
\overline{\gamma_{5}} 
\nonumber \\
& {\rm and }& \nonumber \\
 \overline{\gamma_{3}} =  \overline{\gamma_{6}} = \overline{\gamma_{7}} &=& 
\overline{\gamma_8}~, 
\label{decpars}
\end{eqnarray}
For the decoherence parameters we have chosen (c.f. (\ref{decpars})) 
\begin{equation} 
\overline{\gamma_{1}} = 2 \cdot 10^{-18} \cdot E ~~{\rm and}~~ 
\overline{\gamma_{3}} = 1 \cdot 10^{-24}/E~, 
\label{decohparam}
\end{equation}
In the above formulae
$E$ is  the  neutrino energy, 
and barred quantities
refer to the antineutrinos. 
This parametrisation guarantees positivity 
of the relevant probabilities.
Overall, we have introduced only
two new parameters, two new degrees of freedom, $ \overline{\gamma_{1}}$
and $ \overline{\gamma_{3}}$, which, as argued in \cite{bm} 
was sufficient to account for the available experimental data,
including the ``anomalous'' LSND results.
Furthermore, we have also set the 
CP violating phase of the NMS matrix to zero, so that all the mixing
matrix elements become real.

Since the neutrino sector does not suffer from
decoherence, there is no need to include the solar data into the fit.
We are guaranteed to have an excellent agreement with solar data, as long
as we keep the relevant mass difference and mixing angle within
the LMA region, something which we shall certainly do.

As mentioned previously, CPT violation is driven by, and restricted to, the
decoherence parameters, and hence masses and mixing angles are
the same in both sectors, and selected to be 

\centerline{$\Delta m_{12}^2 = \Delta \overline{ m_{12}}^2 = 
7 \cdot 10^{-5}$~eV$^2$,}
\centerline{$\Delta m_{23}^2 = \Delta \overline{ m_{23}}^2 = 
2.5 \cdot 10^{-3}$~eV$^2$,}
\centerline{$\theta_{23} = \overline{\theta_{23}}= \pi/4$, $\theta_{12} = 
\overline{\theta_{12}}= .45$,}
\centerline{$\theta_{13} = \overline{\theta_{13}}= .05$,} 
\noindent 
as indicated by the state of the art analysis.

At this point it is important to stress that the inclusion of two 
new degrees of freedom 
is not sufficient to guarantee that one will indeed be able to
account for all the experimental observations.
We have
to keep in mind that, in no-decoherence
situations, the addition of a sterile neutrino (which comes
along with four new degrees of freedom -excluding again the possibility
of CP violating phases) did not seem to be sufficient for 
matching 
all
the available experimental data, at least in CPT conserving situations.

In order to test our model 
with these two decoherence parameters in the antineutrino sector,  
we have calculated the zenith angle dependence of the 
ratio ``observed/(expected in the no oscillation case)'', for muon and
electron atmospheric neutrinos, for the sub-GeV and multi-GeV energy ranges,
when mixing is taken into account. Since matter effects are important for
atmospheric neutrinos, we have implemented them through a 
two-shell model, where 
the density in the mantle (core) is taken to be roughly 3.35 (8.44) gr/cm$^3$,
and the core radius is taken to be 2887 km. 
We should note at this stage that a ``fake'' CPT Violation 
appears due to matter
effects, arising from a relative sign difference of the matter
potential between the respective interactions of 
neutrinos and antineutrinos with ordinary matter. This, however, 
is easily disentangled from our genuine (due to quantum gravity)
CPT Violation, used here to parametrise our model fit to LSND results; 
indeed, a systematic study 
of such effects~\cite{ohlsson} has shown that ``fake'' CPT
Violation increases with the oscillation length, but decreases
with the neutrino energy, $E$, vanishing in the limit $E \to \infty$;  
moreover, no independent information
regarding such effects 
can be obtained by looking at the antineutrino sector, as
compared with data from the neutrino sector,  
due to the fact that in the presence of ``fake'' CPT Violation, 
but in the absence of any genuine CPT breaking, 
the pertinent CPT probability differences between 
neutrinos and antineutrinos are related, 
$\Delta P^{\rm CPT}_{\alpha\beta} = 
-\Delta P^{\rm CPT}_{\overline{\beta}\overline{\alpha}}$, 
where $\Delta P_{\alpha\beta}^{\rm CPT}=P_{\alpha\beta}-P_{\overline \beta \overline \alpha}$, and the Greek indices denote neutrino flavors.
These
features are to be contrasted with our dominant decoherence
effects $\overline \gamma_1$ (\ref{decohparam}), 
proportional to the antineutrino energy, $E$, which are dominant 
only in the antineutrino sector. For the same reason, 
our effects can be disentangled from ``fake'' decoherence 
effects arising from Gaussian averages of the oscillation 
probability due to, say, uncertainties in the energy 
of the neutrino beams~\cite{ohlsson2}, which are the same
for both neutrinos and antineutrinos. We, therefore, claim that the
complex energy dependence in (\ref{decohparam}), 
with {\it both} $L\cdot E$ and $L/E$ terms being present in the 
antineutrino sector, 
may be a characteristic  feature of new physics, with the $L \cdot E$ 
terms being related to quantum-gravity induced 
(genuine) CPT Violating decoherence. 

The results are shown in Fig. 1 (c),  
where, for the sake of comparison,
we have also included 
the experimental data.  
We also present in that figure the pure decoherence scenario in the
antineutrino sector (a), as well as in both sectors (b).  
For completeness, we also present a scenario with neutrino mixing
but with 
decoherence 
operative in both sectors (d).  The conclusion is straightforward:
pure decoherence is wildly excluded, while decoherence plus mixing
provides an astonishing agreement with experiment.

\begin{figure*}
\includegraphics[width=8cm]{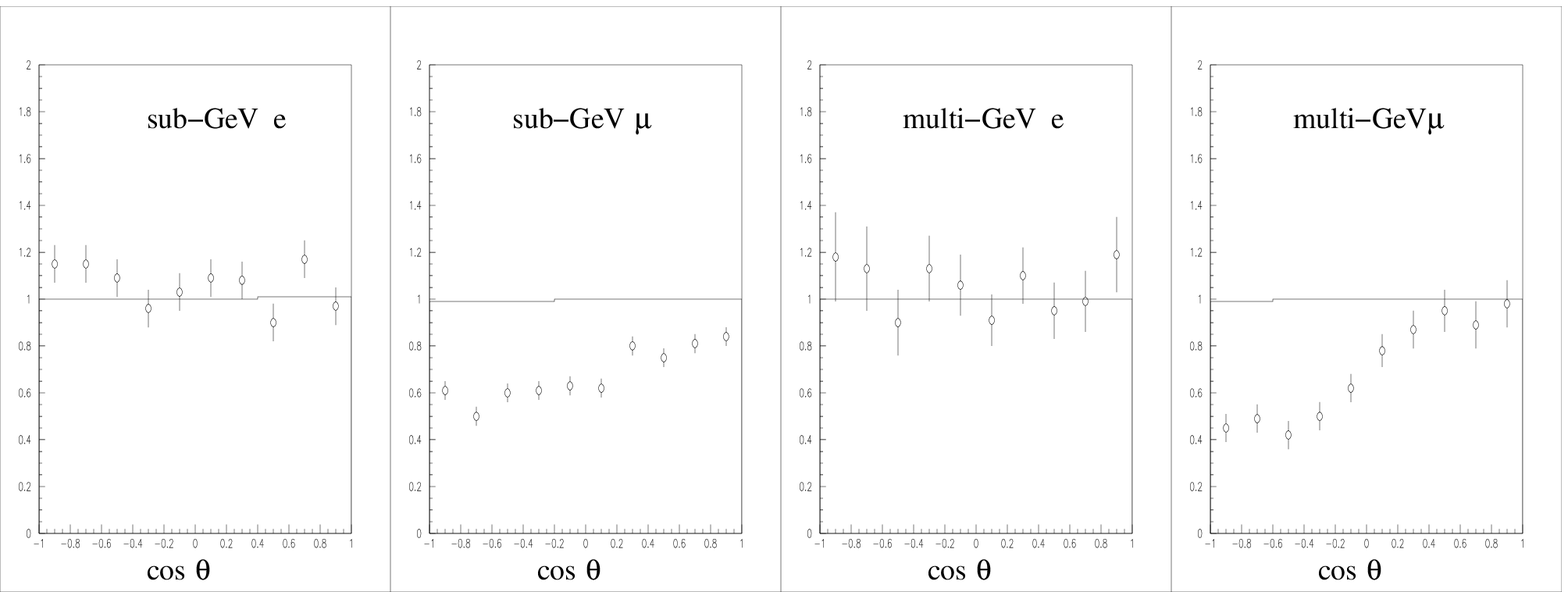} \hfill 
\includegraphics[width=8cm]{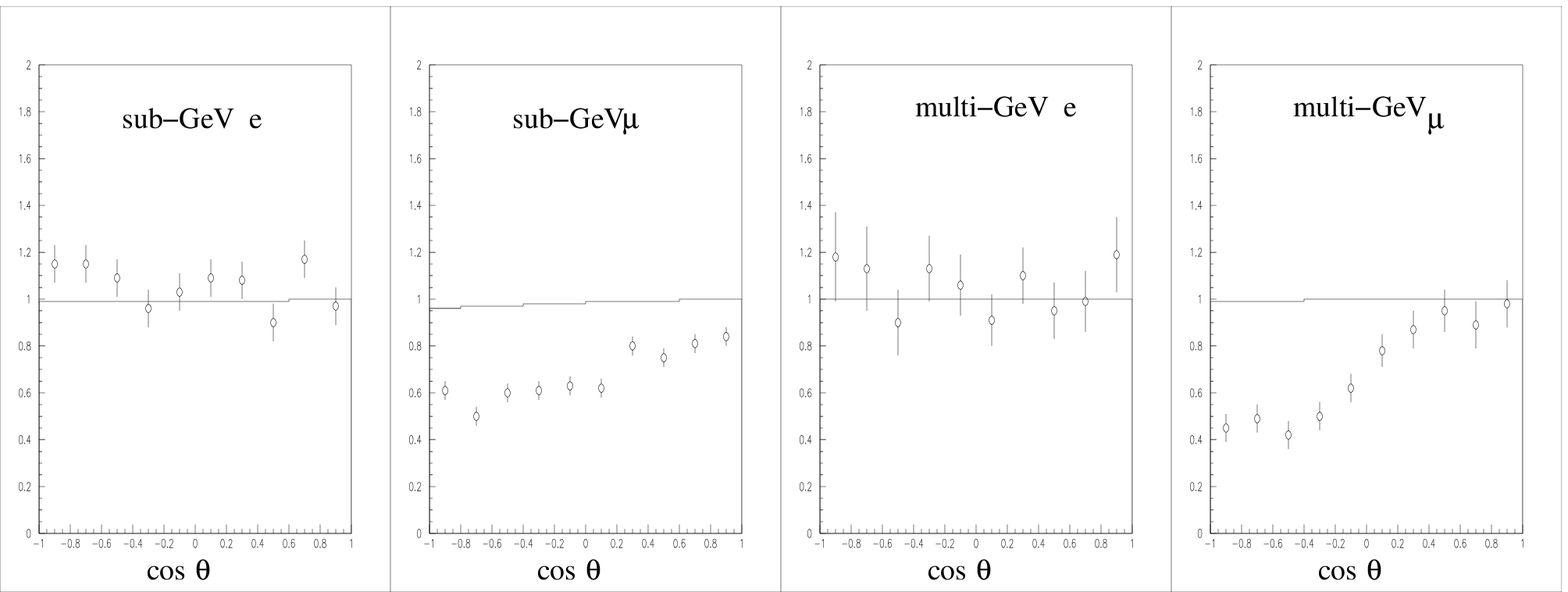} \hfill
\includegraphics[width=8cm]{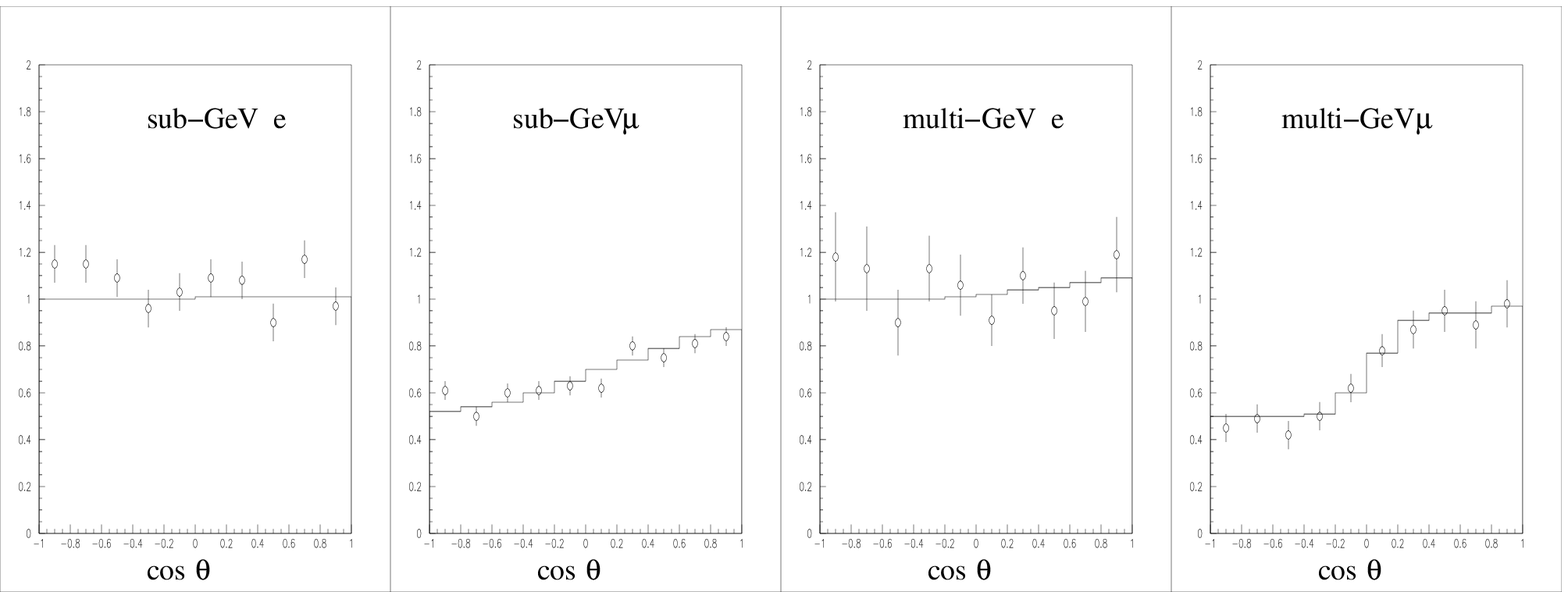}\hfill
\includegraphics[width=8cm]{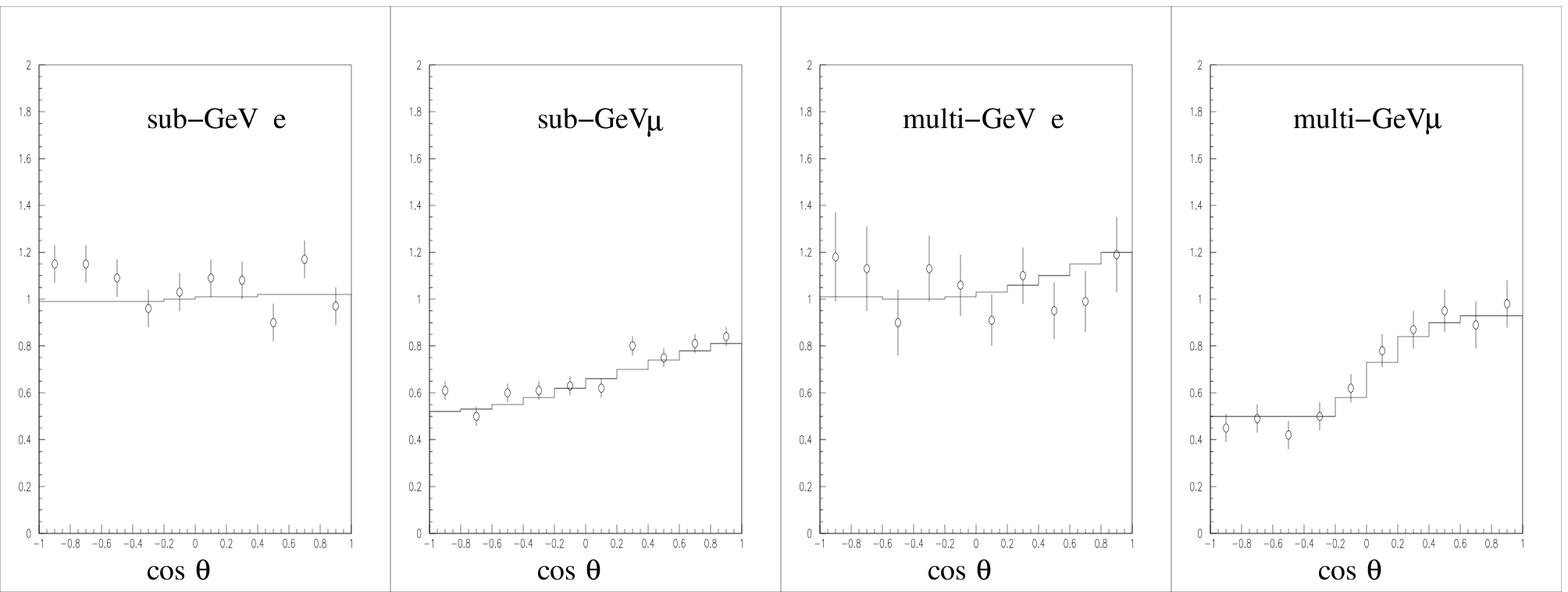}
\caption{Decoherence fits, from top left to bottom right: 
(a) pure decoherence in 
antineutrino sector, (b) pure decoherence in both sectors, (c) mixing plus
decoherence in the antineutrino sector only, (d) mixing plus decoherence
in both sectors. The dots correspond to SK data.}
\label{bestfit}
\end{figure*}

As bare eye comparisons can be misleading, we have also calculated the
$\chi^2$  value for each of the cases~\cite{bm}.From 
this analysis it becomes clear that the mixing plus decoherence scenario
in the antineutrino sector can easily account for all the available 
experimental information,
including LSND. 
It is important to stress once more that our sample
point was not obtained through a scan over all the parameter space,
but  by an educated guess, and therefore plenty of room is left
for improvements. On the other hand, for the mixing-only/no-decoherence 
scenario,
we have taken the best fit values of the state of the art analysis
and therefore no significant improvements are expected. 
At this point a word of warning is in order: although superficially
it seems that scenario (d), decoherence plus mixing in both sectors,
provides an equally good fit, one should remember that including
decoherence effects in the neutrino sector can have undesirable
effects in solar neutrinos, especially due to 
the fact that decoherence
effects are weighted by the distance travelled by the neutrino, 
something that may lead to seisable (not observed!) effects in the
solar case.

One might wonder then, whether decohering effects, which affect
the antineutrino sector sufficiently to account for the LSND result, 
have any impact on the solar-neutrino 
related parameters, measured through 
antineutrinos in the KamLAND experiment \cite{kamland}. 
In order to answer this question,
 it will be sufficient to calculate the electron survival probability
 for KamLAND in our model, which turns out to be 
$ P_{\bar\nu_{\alpha}\rightarrow \bar\nu_{\beta}} \mid_{\mbox{\tiny  KamLAND}}
\simeq .63$, in perfect agreement with observations. As is well known, 
KamLAND is sensitive to a bunch of different 
reactors with distances spanning from 80 to 800 km. However, the bulk of the
signal comes from just two of those, whose distances are 160 and 179 km. These
parameters have been used to compute the survival probability.
It is also interesting
to notice that in our model,
  the LSND effect is not given  by the phase inside
the oscillation term ( which is proportional to the solar mass difference)
 but rather by the decoherence factor multiplying the oscillation term.
 Therefore the tension between LSND and
KARMEN  \cite{karmen} data is naturally eliminated, because the difference
in length leads to an exponential suppression.
Another potential source of concern for the present model of 
decoherence 
might be accelerator neutrino
experiments, which involve high energies and long baselines, and where
the decoherence $L \cdot E$ scaling can potentially be probed. 
This, however, 
is not the case. Accelerator experiments typically join their neutrino and
antineutrino data, with the antineutrino statistics being always smaller than
the neutrino one. This fact, together with the smaller antineutrino cross
section, renders our potential signal consistent with the background
contamination. Even more,  in order to constrain decoherence
effects of the kind we are proposing here through accelerator experiments,
excellent control and knowledge of the beam background are mandatory. 
The new KTeV  data \cite{kk} on kaon decay branching 
ratios, for example, will change the $\nu_e$ background enough to make any 
conclusion on the viability of decoherence models useless. After all, the 
predicted signal  in our decoherence scenario will be at the level of the 
electron neutrino contamination, and therefore one would need to disentangle 
one from the other.

Having said that,
it is now clear that decoherence models (once neutrino mixing is taken
into account) are the best (and arguably the only) way to explain 
all the observations including 
the LSND result. This scenario , which makes dramatic predictions for
the upcoming neutrino experiments, expresses a strong observable form of 
CPT violation in the 
laboratory, and in this sense, our fit gives a clear answer to the 
question asked in the introduction as to whether the weak form of CPT 
invariance (\ref{equality}) is violated in Nature. It seems that, in order to 
account for the LSND results, we should invoke such a decoherence-induced 
CPT violation, which however is independent of any mass differences
between particles and antiparticles.

This CPT violating pattern, with equal mass spectra for
neutrinos and antineutrinos, will have dramatic signatures in 
future neutrino oscillation experiments. The most striking 
consequence
will be seen in MiniBooNE~\cite{miniboone},
According to our picture, MiniBooNE will be able to confirm LSND
only when running in
the antineutrino mode and not in the neutrino one, as decoherence
effects live only in the former. Smaller but experimentally 
accessible signatures
will be seen also in MINOS~\cite{minos}, by comparing conjugated channels (most
noticeably, the muon survival probability).
We should mention at this stage that our model is in 
agreement with the strong suppression
of decoherence in the neutrino sector expected from 
astrophysical observations of high energy cosmic neutrinos~\cite{winst}. 

We next remark that fits with decoherence
parameters with energy dependences of the form (\ref{decohparam}) 
imply that the 
exponential factors $e^{\lambda_k t}$ in (\ref{trans})
due to decoherence will  modify the amplitudes of the oscillatory terms
due to mass differences, and while one term  depends on  
$L/E$ the other one is driven by $L\cdot E$, where
we have set $t=L$, with $L$ the oscillation 
length (we are working with natural units where $c=1$). 

The order of the coefficients of these quantities, 
$\gamma^0_j  \sim 10^{-18}, 10^{-24}$~(GeV)$^2$, found in 
our sample point, implies that for energies of a few GeV,
which are typical of
the pertinent experiments, 
such values are not far from  
$\gamma_j^0 \sim \Delta m_{ij}^2$. If our conclusions
survive the next round of experiments, and therefore if 
MiniBOONE experiment~\cite{miniboone} confirms previous LSND claims,
then this may be a significant result. One would be tempted to 
conclude that if the above estimate holds,  
this would probably mean that 
the neutrino mass differences
might 
be due to quantum gravity decoherence. 
Theoretically it is still 
unknown how the neutrinos acquire a mass, or what kind of mass 
(Majorana
or Dirac) they possess. There are scenaria in which the mass of neutrino 
may be due to some peculiar backgrounds of string theory for instance.
If the above model turns out to be right we might then have, 
for the first time in low energy physics, 
an indication of a direct detection of a quantum gravity effect, which 
disguised itself as an induced decohering neutrino mass difference. 
Notice that in our  sample point only antineutrinos have non-trivial 
decoherence parameters $\overline{\gamma_{i}}$ , for $i=1$ and 3,
while the corresponding quantities in the neutrino sector vanish.  
This implies that there is a single cause for mass differences,
the decoherence in antineutrino sector, which is compatible 
with common mass differences in both  sectors. 
If this turns out to be true, it could then lead to important
conceptual changes in our thinking of the problem of particle
masses in field theory. 

\section{Neutrino Mixing, Space-Time Foam and Cosmological Constant?} 

In what follows  we  will make this assumption, namely that decoherence 
effects, due to interactions with the foam, contribute 
to the Hamiltonian terms in the evolution of the neutrino density 
matrix, and result in  
neutrino mass differences in much the same way as the celebrated
MSW effect\cite{msw}, responsible for a neutrino mass 
splitting due 
to interactions  with a medium. 
Indeed, when 
neutrinos travel through matter, the 
neutral current contribution to this interaction, proportional 
to -$G_F n_n/\sqrt{2}$, with $G_F$ Fermi's weak interaction constant, 
and $n_n$ the neutron density in the medium, 
is present for {\it both} $\nu_e$ and 
$\nu_\mu$ (in a two flavour scenario), while the charged current contribution, given by 
$\sqrt{2}G_Fn_e$, with $n_e$ the medium's electronic density, 
is present only for $\nu_e$. The flavour eigenstates 
$\nu_{e,\mu}$ can then be expressed in terms of fields
${\tilde \nu}_{1,2}$ with definite masses ${\tilde m}_{1,2}$ 
respectively, with a mixing angle ${\tilde \theta}$, the tilde notation
indicating the effects of matter. The tilded quantities are diagonalised
with respect to the Hamiltonian of $\nu_e$,$\nu_\mu$ in the presence 
of non-trivial matter media, and one can find the 
following relations between vacuum (untilded) and medium parameters\cite{msw}
${\rm sin}^22{\tilde \theta} \simeq {\rm sin}^22\theta 
\left(\frac{\Delta m^2}{\Delta {\tilde m}^2}\right)$, with  
$\Delta {\tilde m}^2 = \sqrt{(D -\Delta m^2{\rm cos}2\theta )^2 +
(\Delta m^2{\rm sin}2\theta)^2}, D=2\sqrt{2}G_Fn_e k$. From this 
we observe that the medium-induced effects in the 
mass splittings are proportional 
to the electronic density of the medium and in fact, even if the neutrinos
would have been mass degenerate in vacuum, such a degeneracy would 
be lifted by 
a medium.

To get a qualitative idea of what might happen with 
the foam, one imagines a similar mixing for neutrinos, as a result 
of their interaction with a quantum-gravity decohering foam situation.
As a result, there are {\it gravitationally-induced effective masses} 
for neutrinos,
due to flavour dependent interactions of the foam, which are in principle
allowed in quantum gravity. In analogy (but we stress that 
this is only an analogy) 
with the MSW effect, the gravitationally-induced 
mass-splitting effects are expected now to be proportional
to $G_N n_{\rm bh} k$,
where $G_N =1/M_P^2$ is Newton's constant,
$M_P \sim 10^{19}$ GeV is the quantum gravity scale, 
and $n_{bh}$ is a ``foam'' density of appropriate 
space time defects (such as Planck
size black holes {\it etc.}), whose interaction with the neutrinos
discriminates between flavours, in an analogous way 
to the matter effect. Neutrinos, being 
electrically neutral can indeed interact 
non-trivially with a space time foam, and change flavour as a result
of such interactions, since such processes are allowed by quantum gravity. 
On the other hand, due to electric
charge conservation of microscopic black holes, 
quarks and charged leptons, 
cannot interact non-trivially with the foam. 
In this spirit, one can imagine a microscopic 
charged black-hole/anti-black-hole pair being 
created by the foam vacuum.
Evaporation of these black holes 
(probably at a slower rate than their neutral counterparts, due to
their near extremal nature~\cite{ebh})
can produce 
preferentially $e^+e^-$ pairs (lighter than muons), 
of which the positrons, say, are absorbed into the microscopic
event horizons of the evaporating 
charged anti-black hole.
This leaves us with a stochastically fluctuating (about a mean value)
electron (or more general charge) density, $n^c_{\rm bh}(r)$, 
induced by the gravitational foam, 
$\langle n^c_{\rm bh} (r)\rangle = n_0 \ne 0$, 
$\langle n^c_{\rm bh} (r)n^c_{\rm bh} (r')\rangle \ne 0$,
which, in analogy
with the electrons of the MSW effect in a stochastically
fluctuating medium\cite{loreti}, can interact non-trivially only
with $\nu_e$ but not with the $\nu_\mu$, in contrast to 
neutral
black holes which can interact with all types of neutrinos\cite{bm2}. 
We assume, of course, that the 
contributions 
to the  
vacuum energy that 
may result from such 
emission and absorption processes
by the black holes in the foamy vacuum  
are well within the known limits. For instance,
one may envisage supersymmetric/superstring models of space-time foam, 
where such contributions may be vanishingly small\cite{west}.
The mean value (macroscopic) part, $n_0$, of 
$n^c_{\rm bh}(r)$, assumed time independent, will contribute to the 
Hamiltonian part of the evolution of the neutrino density matrix, $\rho$.   
In analogy with the (stochastic) MSW effect\cite{msw,loreti},
this part yields space-time foam-induced 
mass-squared splittings for neutrinos:
\begin{equation}
\langle\Delta m^2_{\rm foam}\rangle 
\propto G_N \langle n^c_{\rm bh}(r)\rangle k
\label{mswgrav}
\end{equation}
with non trivial quantum fluctuations ($k$ is the neutrino momentum scale).  
To ensure a constant neutrino mass one may consider the case where 
$\langle n^c_{\rm bh}(r)\rangle$, which expresses the average number 
of virtual particles emitted from the foam with which the neutrino
interacts, is inversely proportional to 
the (neutrino) momentum. This is reasonable, since the faster the neutrino, 
the less the available time to interact with the foam, and hence 
the smaller the number of foam particles 
it interacts with. Such flavour-violating foam 
effects would also contribute to decoherence
through the quantum fluctuations of the foam-medium 
density\cite{loreti,bm2},
by means of induced non-Hamiltonian 
terms in the density-matrix evolution. 
Such effects assume a double commutator structure\cite{loreti,bm2,adler}
and are due to {\it both}, the fluctuating parts of the foam density, as well
as the effects of the mixing (\ref{mswgrav}) 
on the vacuum energy. 
Indeed, as we discussed in \cite{bm2}, and shall review briefly below, 
neutrino flavour mixing may lead
to a non-trivial contribution to the vacuum energy, in a non-perturbative 
way suggested in \cite{vitiello2}.
Hence, 
such effects are necessarily CPT violating\cite{mlambda}, 
in the sense of entailing an evolution 
of an initially pure neutrino quantum state to a 
mixed one due to the presence of the Hubble horizon
associated with the non zero cosmological constant,
which prevents pure asymptotic states from being well defined. 
In that case, CPT is violated  
in its strong form, that is CPT is not a well-defined operator, 
according to the theorem of \cite{wald}.

For convenience we shall discuss explicitly the two-generation
case. The arguments can be extended to three generations, 
at the expense of an increase in mathematical complexity,
but will not affect qualitatively the conclusions
drawn from the two-generation case. 
The arguments are based on the observation\cite{vitiello1} 
that in quantum field theory, which by definition requires an 
infinite volume limit, in contrast to 
quantum mechanical treatment
of fixed volume\cite{pontecor}, the neutrino {\it flavour} states are 
{\it orthogonal} to the 
{\it energy} eigenstates, and moreover they define 
two inequivalent vacua
related to each other by a {\it non unitary} transformation $G^{-1}(\theta,t)$:
$|0(t)\rangle_f = G^{-1}_\theta (t)|0(t)\rangle_m$,
where $\theta$ is the mixing angle, $t$ is the time, and the suffix f(m)
denotes flavour(energy) eigenstates respectively, and 
$G^{-1}_\theta (t) \ne G^{\dagger}_\theta (t)$ is a non-unitary operator
expressed in terms of energy-eigenstate neutrino free fields $\nu_{1,2}$\cite{vitiello2}: $G_\theta (t) = {\rm exp}\left(\theta \int d^3x [\nu_1^\dagger (x)
\nu_2 (x) - \nu_2^\dagger (x)
\nu_1(x)]\right)$.
A rigorous mathematical
analysis of this problem has also appeared in \cite{hannabus}. 
As a result of the non unitarity of $G^{-1}_\theta (t)$, there is a 
Bogolubov transformation\cite{vitiello1} 
connecting the creation and annihilation operator 
coefficients appearing in the expansion of the appropriate  
neutrino fields of the energy or flavour eigenstates. 
Of the two Bogolubov coefficients appearing in the
treatment, we shall concentrate on $V_{\vec k} = 
|V_{\vec k}|e^{i(\omega_{k,1} +\omega_{k,2})t} $, with 
$\omega_{k,i}=\sqrt{k^2 + m_i^2}$,  
the (positive) energy of the neutrino energy eigenstate $i=1,2$ 
with mass $m_i$. This function is related to the condensate 
content of the flavour vacuum, in the sense 
of appearing in the expression of an appropriate 
non-zero number operator of the flavour vacuum\cite{vitiello2,hannabus}:
$_f\langle 0 |\alpha_{{\vec k}, i}^{r \dagger}\alpha_{{\vec k}, i}^{r}
|0\rangle_f = _f\langle 0 |\beta_{{\vec k}, i}^{r \dagger}
\beta_{{\vec k}, i}^{r}|0\rangle_f = {\rm sin}^2\theta |V_{\vec k}|^2$ 
in the two-generation scenario~\cite{vitiello1}. 
$|V_{\vec k}|$ has the property of vanishing for $m_1=m_2$, it has a maximum 
at the momentum scale $k^2 =m_1m_2$, and for $k \gg \sqrt{m_1m_2}$ 
it goes to zero as:
\begin{equation}
|V_{\vec k}|^2 \sim \frac{(m_1 - m_2)^2}{4|{\vec k}|^2}, \quad 
k \equiv |{\vec k}| \gg \sqrt{m_1m_2}
\label{bogolubov}
\end{equation}
The analysis of \cite{vitiello2} argued that the flavour vacuum $|0\rangle$, 
is the
correct one to be used in the calculation of the average vacuum energy, 
since otherwise the probability is not conserved\cite{henning}.
The energy-momentum tensor of a Dirac fermion field in the Robertson-Walker
space-time background can be calculated straightforwardly in this formalism. 
The flavour-vacuum average value of its temporal $T_{00}$ component,
which yields the required contribution to the vacuum energy 
due to neutrino mixing, is\cite{vitiello2}:
\begin{eqnarray}
&& _f\langle 0|T_{00} |0\rangle_f 
=\langle \rho_{\rm vac}^{\rm \nu-mix}\rangle \eta_{00} \nonumber \\
&& = \sum_{i,r}\int d^3 k \omega_{k,i}\left(_f\langle 0 |\alpha_{{\vec k}, i}^{r \dagger}\alpha_{{\vec k}, i}^{r}|0\rangle_f + 
_f\langle 0 |\beta_{{\vec k}, i}^{r \dagger}\beta_{{\vec k}, i}^{r}|0\rangle_f\right) 
= \nonumber \\
&& 8{\rm sin}^2\theta \int_0^K d^3k (\omega_{k,1} +  
\omega_{k,2})|V_{\vec k}|^2.
\label{vacener}
\end{eqnarray}
where $\eta_{00}=1$ in a Robertson-Walker (cosmological) metric background.
The momentum integral in (\ref{vacener}) is cut-off 
from above at a certain scale, $K$ 
relevant to the physics of neutrino mixing. In conventional approaches,
where the mass generation of neutrino occurs at the electroweak phase 
transition, this cutoff scale can be put on the electroweak
scale $K \sim 100 $ GeV, but this yields unacceptably large contributions to
the vacuum energy. An alternative scale has been suggested 
in \cite{vitiello2}, namely $K \sim \sqrt{m_1m_2}$ as the
characteristic scale for the mixing. 
In this way these authors 
obtained a phenomenologically acceptable value
for $\langle \rho_{\rm vac}^{\rm \nu-mix} \rangle $. 

In our case we 
shall use a different cutoff scale~\cite{bm2}, 
which allows for some analytic estimates of (\ref{vacener})
to be derived, as being  
mathematically consistent with the 
asymptotic form of (\ref{bogolubov}), which  
is valid in a regime of momenta $k \gg \sqrt{m_1m_2}$.
This cutoff scale is simply given by the sum of 
the two neutrino masses, 
$K \equiv k_0 = m_1 + m_2$, 
is compatible with our decoherence-induced mass difference scenario,
and also allows for a mathematically consistent analytic estimate of the 
neutrino-mixing contribution to the vacuum energy in this framework.
For hierarchical neutrino models, for which $m_1 \gg m_2$, we have 
that $k_0 \gg \sqrt{m_1m_2}$, and thus, if we assume that the 
modes near the cutoff contribute most to the vacuum energy 
(\ref{vacener}), which is clearly supported by the otherwise divergent 
nature of the momentum integration, and take into account the 
asymptotic properties of the function $V_{\vec k}$, which are safely valid
in this case, we obtain: 
\begin{eqnarray} 
&&\langle \rho_{\rm vac}^{\rm \nu-mix}\rangle \sim 
8\pi{\rm sin}^2\theta (m_1 - m_2)^2 (m_1 + m_2)^2\times \nonumber \\ 
&&\left(\sqrt{2} + 1 +{\cal O}(\frac{m_2^2}{m_1^2})\right) \propto {\rm sin^2}\theta (\Delta m^2)^2
\label{darkneutr}
\end{eqnarray}
in the limit $m_2 \ll m_1$. 
For the (1,2) sector, the corresponding $\Delta m^2 $ is given by the solar neutrino
data and is estimated to be $\Delta m^2_{12} \simeq 10^{-5}$ eV$^2$, resulting in a
contribution of the right order.
In this way 
the cosmological constant $\Lambda$ is  
elegantly expressed in terms 
of the smallest (infrared, $\Delta m^2$) and the largest 
(ultraviolet, $M_P^2$) Lorentz-invariant mass scales available.
It can be argued~\cite{bm2} that the above 
choice of the cutoff $k_0 \sim m_1 + m_2$ is 
consistent with our conjecture on the decoherence origin of the neutrino
mass difference, due to interaction with the foam medium (\ref{mswgrav}).
Notice that the above way of deriving the neutrino-mixing contribution 
to the dark energy is independent of the usual perturbative loop arguments,
and, in this sense, the result (\ref{darkneutr}) 
should be considered as  exact 
(non perturbative), if true.

Some important remarks are now in order. 
First of all, our choice of cutoff scale was such that the 
resulting contribution to the cosmological constant depends on the 
neutrino mass-squared differences and not on the absolute mass,
and hence it is independent of any zero-point energy, in agreement
with energy-driven decoherence models~\cite{adler}.
For us, it is curved space physics that is  responsible for 
lifting the mass degeneracy of neutrino mass eigenstates and create the
``flavour'' problem. 
This is an important point, which may serve as motivation (not proof)
behind such a cutoff ``choice'', which we conjecture 
is a physical ``necessity''.
We have argued above that such a cutoff ``choice'' 
is a natural one from the point of 
view of quantum-gravity decoherence-induced mass differences. 
Detailed models of this fall way beyond the purposes of this brief note.
Nevertheless, we believe 
that the above-demonstrated self-consistency of this cutoff choice
within the remit  
of our toy model of space time foam 
is intellectually challenging and encouraging for further studies
of this important issue. 

The  above 
considerations above were based on the suggestion 
of ref. \cite{vitiello1} on a Fock-like quantisation of 
the flavour space. There is still controversy in the literature 
regarding the physical meaning of such quantum flavour states~\cite{giunti},
in particular it has been argued that, although such states are mathematically
elegant and correct constructions, 
nevertheless they lead to no observable consequences. 
However, 
in view of the results of \cite{vitiello2} and of the present work,
such an argument may not be correct, since the 
mass-squared difference contribution to the cosmological constant
is an observable (global) consequence of the Fock-like flavour space 
quantisation.
The presence of a time independent cosmological constant (\ref{darkneutr})
in the flavour vacuum, which notably is not present if one uses instead the
mass eigenstate vacuum, 
implies an asymptotic future event horizon for the emerging 
de-Sitter Universe. 
The flat-space time arguments of 
\cite{giunti} for the flavour space field theory 
cannot then be applied, at least naively, and the problem 
of quantisation
of the Fock-like flavour space is equivalent to the (still elusive) 
quantisation of field theories in (curved) de-Sitter space times.
In such a case one cannot define properly 
asymptotic states, and hence a scattering matrix. 
This will lead to decoherence, in the sense of a modified temporal 
evolution for matter states. 

We now remark that in the case of 
(anti)neutrinos passing through stochastic media~\cite{loreti}, including 
space time foam~\cite{bm2},  
there are additional contributions to decoherence, besides the presence of a 
$\Lambda$-term,  
which may offer
a natural explanation of the decoherence parameters of \cite{bm}.
An important source of decoherence in such media is due to the 
uncertainties in the energy $E$ and/or the oscillation 
length $L$ of the (anti)neutrino beam. 
In fact, it can be shown~\cite{ohlsson2}
that if one averages the standard oscillation 
probabilities $P_{\nu_\alpha \to \nu_\beta}$
over Gaussian distributions for $E$ and/or $L$ with a variance $\sigma^2$,
the result is equivalent to neutrino decoherence models, in the sense
of the time dependent profile of the associated 
probability being identical to that of a completely-positive  
decoherence model. One finds for $n$ flavours~\cite{ohlsson}, 
\begin{eqnarray} 
&& \langle P_{\alpha\beta} \rangle = \delta_{\alpha\beta} 
- 
2 \sum_{a=1}^n\sum_{\beta=1, a<b}^n{\rm Re}\left(U_{\alpha a}^*
U_{\beta a}U_{\alpha b}U_{\beta b}^*\right)\times \nonumber \\
&& \left( 1 - {\rm cos}(2\ell \Delta m_{ab}^2)
e^{-2\sigma^2(\Delta m_{ab}^2)^2}\right) -
\nonumber \\
&& 2 \sum_{a=1}^n \sum_{b=1, a<b}^n {\rm Im}\left(U_{\alpha a}^*
U_{\beta a}U_{\alpha b}U_{\beta b}^*\right) \times \nonumber \\
&& {\rm sin}(2\ell \Delta m_{ab}^2)
e^{-2\sigma^2(\Delta m_{ab}^2)^2}
\label{avprobs}
\end{eqnarray}
where $U$ is the mixing matrix 
$\ell \equiv \langle x \rangle$, 
$\sigma = \sqrt{\langle (x - \langle x \rangle)^2} \equiv (L/4E)r$, 
and $x = L/4E$.
The resulting form is identical to that of decoherence, as becomes evident
by noting that the  
exponential damping factors can be written 
in the form $e^{-\gamma_j L}$
with $t=L~(c=1)$, and 
decoherence parameters $\gamma_j$ of order: 
$2\sigma^2_j (\Delta m^2)^2 = \gamma_j L$, from which  
$\gamma_j = \frac{(\Delta m^2)^2}{8E^2}Lr^2_j$.
There are various 
scenaria that restrict the order of $\sigma$.
In general, the
acceptable bounds on $\sigma$ may be divided in two 
major categories, depending on the form of the uncertainties~\cite{ohlsson2}:
$\sigma_j \simeq \Delta x \simeq \Delta_j\frac{L}{4E}
\le \frac{\langle L \rangle}{4 \langle E \rangle}
\left(\frac{\Delta_j L}{\langle L \rangle} 
+ \frac{\Delta_j E}{\langle E \rangle}\right)$, or 
$\sigma_j 
\le \frac{\langle L \rangle}{4 \langle E \rangle}
\left([\frac{\Delta_j L}{\langle L \rangle}]^2 + 
[\frac{\Delta_j E}{\langle E \rangle}]^2\right)^{1/2}$. 
In three generation models the values of the length and energy uncertainties
may vary between flavours, and also between neutrinos and antineutrinos,
as a result of the intrinsic CPT violation, hence the subscript $j$ 
in the above formulae (for 
antiparticle sectors it is understood that $j \to {\overline j}$).
From the above considerations it becomes clear that, for 
$L \sim 2E/\Delta m^2$, which is characteristic for oscillations, 
one has decoherence parameters $\gamma_j \sim (\Delta m^2/E)r_j^2$.
It is interesting to estimate first the order of 
decoherence induced by conventional physics, 
for instance decoherence induced by uncertainties
in the measured energy of the beam due to experimental limitations.
For long base line, atmospheric or cosmic neutrino experiments,  
where $\Delta L/L$ is negligible, and $\Delta E/E \sim 1$ 
such decoherence 
parameters are found 
at most of order $\gamma \sim 10^{-24}$ GeV, 
for the relevant range of energies, 
and they diminish with 
energy, vanishing formally when $E \to \infty$, which seems to be a 
general feature of conventional matter-induced  decoherence 
effects~\cite{ohlsson}. 
 
To obtain the decoherence parameters of the best-fit model of 
\cite{bm} it suffices to choose for the antineutrino sector 
$r_{\overline 3}=r_{\overline 8} \sim \Delta E /E \sim 1 $,
and $r_{\overline 1}^2=r_{\overline 2}^2 \sim 10^{-18} \cdot E^2/\Delta m^2 $.
As seen above, the decoherence 
parameters exhibiting a $1/E$ energy dependence
could be attributed 
to conventional energy uncertainties 
occurring in the beam of the (anti)neutrinos. 
However, the parameters proportional to $E$, if true, may be attributed
to exotic physics. The fact that $r_j$ in general receives contributions from 
both length and energy uncertainties provides a natural
explanation for the different energy dependence of the 
decoherence parameter of the model of I in the antiparticle
sector. Indeed, having identified $r_{\overline 3}=r_{\overline 8}$ 
as decoherence 
induced by `conventional-looking'
energy uncertainties in the antineutrino sector, 
it is natural to assume that the $\bar \gamma_1=\bar \gamma_2 \propto E$ 
decoherence
is due to genuine quantum gravity effects, increasing with energy, 
which are associated with metric tensor quantum fluctuations. 
This is achieved provided we assume that 
$r_{\overline 1}^2=r_{\overline 2}^2 \sim (\Delta L/L)^2$,
i.e. these decoherence coefficients are 
predominantly oscillation-length-uncertainty driven,  
and take into 
account that variations in the invariant length may be caused by metric
fluctuations, since $L^2 =g_{\mu\nu}L^\mu L^\nu$, implying
$(\Delta g_{\mu\nu})^2 \sim (\Delta L)^2/L^2$, in order of magnitude.
To obtain the best fit results of I, then, 
for $L \sim 2E/\Delta m^2$, one needs quantum-gravity induced
metric fluctuations in the antineutrino sector 
of order $ (\overline \Delta g_{\mu\nu})^2 \sim 10^{-18} L \cdot E$. 
The increase with energy
is not unreasonable, given that the higher the energy of the antineutrino
the stronger the back reaction onto space time, and hence the 
stronger the quantum-gravity induced metric fluctuations. The factor
$10^{-18}$ may be thought of as being of order $E/M_P$, with
$M_P \sim 10^{19}$ GeV the Planck mass, although 
alternative interpretations may be valid (see 
discussion on possible cosmological interpretations at the end of the 
article). The increase with $L$ is not uncommon in stochastic models
of quantum foam, where the decoherence `medium' 
effects build up with the distance the (anti)particle
travels~\cite{west}.
We also mention at this stage that, 
apart from 
these effects, 
in stochastic models
of foam there are additional contributions to decoherence, arising
from the fluctuations of the density of the medium.
These too can mimic the effects of the best-fit model of \cite{bm}
in the antineutrino sector, as discussed in some detail in \cite{bm2},
but their $L$-dependence is different from that of the above effects.
Comparison between short and long baseline experiments, therefore,
may differentiate between the various decoherent contributions.  

At this stage we would like to mention that 
the above-described 
model of decoherence provides
a novel and extremely economical mechanism to generate
the observed baryon asymmetry in the Universe~\cite{bm2}, 
through a process of
equilibrium electroweak leptogenesis. To this end 
we first recall that, 
in the analysis of ref. \cite{lopez}, 
dealing with decoherent evolution in the neutral Kaon case, 
the asymmetries between
the semileptonic decays of $K_0$ and those of ${\overline K_0}$ 
turned out to depend linearly on dimensionless decoherence 
parameters such as  ${\widehat \gamma} = \gamma/\Delta \Gamma$; 
in the parametrisation of Ellis et al. in \cite{ehns}, 
where $\Delta \Gamma=\Gamma_L - \Gamma_S$ was a characteristic
energy scale associated with energy eigenstates of the kaon system.
In fact, the dependence was such that the decoherence 
corrections to the asymmetry were of order ${\widehat \gamma}$ 
in complete positivity scenaria, where only one decoherence parameter,
$\gamma > 0$ was non zero.
In similar spirit, in our case of lepton-antilepton number asymmetries,
one expects the corresponding asymmetry to depend, to leading order, 
linearly on the quantity 
${\widehat { \gamma}} = \gamma/\sqrt{\Delta 
m^2}$, since the quantity $\sqrt{\Delta m^2}$ is the characteristic 
energy scale in the neutrino case, playing a role analogous to 
$\Delta \Gamma$  in the kaon case. The only difference from the kaon 
case,
is that here, in contrast to the kaon asymmetry results, there are no
zeroth order terms, and thus the result of the matter-antimatter asymmetry
is proportional to the dimensionless decoherence parameter 
$\widehat{\gamma}$, which we are going to take as the larger of the two
dechorence parameters of our model in \cite{bm}, discussed 
in the previous section, namely $
\widehat{\gamma} \to \widehat{\gamma_1} = 10^{-18} \cdot E/\sqrt{\Delta m^2}$.
In this way, the matter-antimatter asymmetry in the 
Universe is estimated to be 
${\cal A} =  \frac{\langle \nu \rangle - \langle {\overline \nu} 
\rangle}{\langle \nu \rangle + \langle {\overline \nu} \rangle}
\simeq  \, \widehat{\gamma_1} \, \simeq  10^{-6}$. The numerical
coefficient $10^{-18}$ on $\gamma$ may be thought of as the 
ratio $T/M_P$ with $T$ the temperature, whose value gets frozen at 
the EW symmetry breaking temperature. 
Thus, $B= \frac{n_\nu -\bar{n_\nu}}{s} \sim 
\frac{{\cal A} n_\nu }{g_* n_\gamma}$ with $n_\nu $ ($\bar{n_\nu} $) 
the number density of (anti) neutrinos, $n_\gamma$ the number density of 
photons and $g_*$ the effective
number of degrees of freedom  (at the temperature where the
asymmetry is developed) which depends on the exact matter content
of the model but it ranges from $10^2$ to $10^3$ in our
case. This implies a residual baryon asymmetry of order $10^{-10}$,
roughly the desired magnitude.  

Finally, before closing we would like to alert the interested
reader in another possible aspect of the Fock space quantisation
of the flavour vacuum~\cite{vitiello2}, advocated in \cite{magueijo}.
In the case of non-trivial neutrino mixing, the flavour Fock states 
do not satisfy the standard Linear Lorentz invariant 
dispersion relations $E^2 = p^2 + m^2$, where $p$ is the momentum
and $m$ a rest mass, since they are superpositions of mass eigenstates
satisfying standard dispersion relations but with different masses. 
The idea of \cite{magueijo} is that such states 
may experience 
non-linear modifications of the Lorentz symmetry, 
of which doubly special relativities is one example~\cite{dsr}, 
which should guarantee the frame independence of the results. 
The remark we would like to make in connection with this, 
is that, in view of the 
contributions of the Flavour Fock states to the cosmological 
constant (\ref{darkneutr}), it may be possible to specify 
the non-linear modifications of the Lorentz symmetry satisfied by the 
Fock states, which was not possible for a general mixing angle 
in \cite{magueijo}, 
by adopting the idea of \cite{smolin}. According to that,  
the low-energy limit of a quantum theory of gravity 
in a space time 
with a (positive) cosmological constant $\Lambda$,
must be a theory which is invariant under a {\it deformed} Poincar\'e 
symmetry, with the pertinent 
(dimensionful) deformation parameter being related to the 
cosmological constant. These arguments are valid 
as long as the theory behaves smoothly
in the limit when the cosmological constant becomes small as compared to the
Planck scale (or, in general the scale characteristic of the 
quantum gravity), {\it i.e.} in the limit $\Lambda \ell_{\rm Planck}^2 \to 0$,
which is certainly the case of (\ref{darkneutr}).      
In our case, where the neutrino mass differences have been conjectured to be
the result of a space-time foamy situation, such considerations 
become of great importance in determining the symmetry
structure underlying the non-flat space-time quantum field (or string) 
theory at hand. We hope to study  such 
important issues in detail in a future publication.

\section{Acknowledgements}

It is a pleasure to thank the organisers of the DICE2004 conference
for the invitation, their support, 
and for creating such a thought stimulating and pleasant
atmosphere during the conference. It is also my 
pleasure to acknowledge discussions 
with M. Blasone, A. Capolupo, G. Lambiase, F. Lombardo, R. Rivers, R. Sorkin 
and G. Vitiello. 

\bibliography{apssamp}

\end{document}